\documentclass[useAMS,usenatbib]{mnras}
\usepackage{amsmath}
\usepackage{mathtools}
\usepackage{amssymb}
\usepackage{graphicx}
\usepackage{multirow}
\usepackage{color}
\usepackage{array}
\usepackage{graphicx}
\usepackage{threeparttable}

\title[X-ray reverberation mapping using the RFR model]{Coronal height constraint in IRAS~13224--3809 and 1H~0707--495 by the random forest regressor}

\author[N. Mankatwit et al.]{N. Mankatwit$^{1}$
, P. Chainakun$^{1,2}$\thanks{E-mail: \href{mailto:pchainakun@g.sut.ac.th}{pchainakun@g.sut.ac.th}}, W. Luangtip$^{3,4}$, A. J. Young$^5$ \\
$^1$School of Physics, Institute of Science, Suranaree University of Technology, Nakhon Ratchasima 30000, Thailand\\
$^2$Centre of Excellence in High Energy Physics and Astrophysics, Suranaree University of Technology, Nakhon Ratchasima 30000, Thailand\\
$^3$Department of Physics, Faculty of Science, Srinakharinwirot University, Bangkok 10110, Thailand\\
$^4$National Astronomical Research Institute of Thailand, Chiang Mai 50180, Thailand\\
$^5$H. H. Wills Physics Laboratory, Tyndall Avenue, Bristol BS8 1TL, UK}
\date{Accepted XXX. Received YYY; in original form ZZZ}

\pubyear{2021}

\begin{document}
\label{firstpage}
\pagerange{\pageref{firstpage}--\pageref{lastpage}}
\maketitle

\begin{abstract}
We develop a random forest regressor (RFR) machine learning model to trace the coronal evolution in two highly variable active galactic nuclei (AGNs) IRAS~13224–3809 and 1H~0707–495 observed with {\it XMM-Newton}, by probing the X-ray reverberation features imprinted on their power spectral density (PSD) profiles. Simulated PSDs in the form of a power-law, with similar frequency range and bins to the observed data, are produced. Then, they are convolved with relativistic disc-response functions from a lamp-post source before being used to train and test the model to predict the coronal height. We remove some bins that are dominated by Poisson noise and find that the model can tolerate the frequency-bin removal up to $\sim 10$ bins to maintain a prediction accuracy of $R^{2} > 0.9$. The black hole mass and inclination should be fixed so that the accuracy in predicting the source height is still $> 0.9$. The accuracy also increases with the reflection fraction. The corona heights for both AGN are then predicted using the RFR model developed from the simulated PSDs whose frequency range and bins are specifically adjusted to match those from each individual observation. The model suggests that their corona varies between $\sim~5 - 18~r_{\rm g}$, with $R^{2} > 0.9$ for all observations. Such high accuracy can still be obtained if the difference between the true mass and the trained value is $\lesssim 10\%$. Finally, the model supports the height-changing corona under the light-bending scenario where the height is correlated to source luminosity in both IRAS~13224–3809 and 1H~0707–495.

\end{abstract}


\section{Introduction}

X-ray reverberation in active galactic nucleus (AGN) occurs when the primary X-ray continuum from the corona illuminates an accretion disc, producing the reflected X-rays that are observed with a time delay after the direct continuum \citep[see][for a review]{Uttley2014, Cackett2021}. The reverberation lags due to the delay between the variations of the reflection and the direct continuum components were detected via the Fourier-based technique \cite[e.g.][]{Fabian2009, Demarco2013, Kara2016}. The lag amplitude depends on the light travel distance between the corona and the disc, providing valuable information about the geometry of the system. Most previous work investigated X-ray reverberation signatures in the lag-frequency spectra using the standard lamp-post geometry where the corona is an isotropic point-like source located on the rotational axis of the black hole \citep{Cackett2014, Emmanoulopoulos2014, Chainakun2016, Epitropakis2016, Caballero2018}, while some investigated it in an extended corona environment \citep{Wilkins2016, Chainakun2017, Chainakun2019b}. 

The AGN emission fluctuates rapidly on a wide range of timescales. The amount of the X-ray variability power that the AGN emits can be quantified using the power spectral density (PSD), whose intrinsic shape can also be imprinted with X-ray reverberation effects  \citep{Papadakis2016,Chainakun2019a}. The observed PSD profiles tend to be relatively flat at low Fourier frequencies, with a steep drop-off at higher frequencies \citep{Martin2012}. The intrinsic shape of the PSD depends on the specific characteristics of the accretion disc \citep{Arevalo2006, Ashton2022}. An attempt at probing the reverberation signature (appeared as an oscillatory structure) in the PSD data of AGN to constrain the coronal height were carried out by \cite{Emmanoulopoulos2016, Chainakun2022}. In this way, the coronal geometry can be constrained, but high accuracy prediction may still require the intrinsic PSD shape to be known in advance.

Furthermore, \cite{Chainakun2021} developed a machine learning (ML) model using dictionary-learning and support vector machine to predict the discrete values of the coronal height. Interestingly, while the reverberation-imprinted and power-law based PSD profiles were used to train the model, it was suggested that the accuracy of predicting the source height was still high even when the model was tested with a bending power-law PSD. Here, we aim to further develop the model so that it can predict a continuous value of the coronal height and can directly be applied to real observational data. We use a random forest regressor (RFR) which is an ensemble ML algorithm that is composed of many individual decision trees being trained on subsets of the data. Each tree makes its individual prediction, and the final prediction is achieved by statistical averaging the predictions of all individual decision trees.

We select to apply our developed RFR model to the AGN IRAS~13224--3809 and 1H~0707--495, both of which are of particular interest due to their unique X-ray spectral properties and extreme variability. IRAS~13224--3809 is a type 1 AGN that is known to have a supermassive black hole at its center, with a mass of $\sim 2 \times 10^6 \ M_{\odot}$ \citep{Alston2020}. It has been thoroughly studied via both spectral and timing analysis \citep[e.g.][]{Fabian2013, Kara2013, Chiang2015, Parker2017,Jiang2018}. Along the line of these studies, the changing of the coronal height with luminosity was evidenced \citep{Alston2020, Caballero2020, Chainakun2022}, where the height increases from $\sim 3$--$5~r_{\rm g}$ to $\sim 10$--$25~r_{\rm g}$ ($r_{\rm g}=GM/c^{2}$, $G$ is the gravitational constant, $M$ is the black hole mass and $c$ is the speed of light) when the source changes from lowest to highest flux state. The model that  simultaneously fitted the energy and lag spectra in various flux states also favoured a maximum spin black hole \citep{Caballero2020}. Meanwhile, the 1H~0707--495 corona geometry is very uncertain and under debate. While an analysis using the emissivity profiles suggested an extended corona for 1H~0707--495 \citep{Wilkins2012, Wilkins2014}, a much more compact corona was obtained using a newly developed spectral model accounting for the spatial extent and the rotation of the corona \citep{Szanecki2020}. Both intrinsic and environmental absorption were also found when analysing the fractional excess variance of 1H~0707--495 \citep{Parker2021}. 

This paper is organised as follows. The observations and data reduction are described in Section~2. How the PSD profiles are produced is explained in Section~3. In Section~4 we present the RFR algorithms used in this work and explain how the RFR model is developed. The results and analysis towards the performance of the model is given in Section~5, as well as the results when we apply the model to predict the coronal height in IRAS~13224--3809 and 1H~0707--495. The discussion and conclusion is provided in Section~6.

\section{Observations and data reduction}

The X-ray data of the AGN IRAS~13224--3809 and 1H~0707--495 used in this study were previously observed by {\it XMM-Newton} observatory \citep{Jansen2001} and were obtained from the {\it XMM-Newton} Science Archive.\footnote{Available online at \url{http://nxsa.esac.esa.int/}} All observational data are shown in Table~\ref{tab:observations}. Here, only pn data are used to obtain high quality data. To analyse the data, we created the PSDs from these observations following the method and criteria explained in \citet{Chainakun2022}. In brief, the pn data were reprocessed using the Science Analysis Software (SAS) following the procedure suggested in {\it XMM-Newton} Science Analysis System's user guide.\footnote{\url{https://xmm-tools.cosmos.esa.int/external/xmm_user_support/documentation/sas_usg/USG.pdf}} The observational periods which were affected by high background flaring activity were also removed during these steps. The remaining useful exposure time of each observation after removing the background flaring events is shown in column 4 of Table~\ref{tab:observations}. Note that, in this step, we only considered the observations with a remaining exposure time of $\ga$30 ks so that the signal to noise of their PSDs is sufficiently high.

\begin{table*}
\begin{center}
   \caption{\emph{XMM-Newton} observations of the AGN sample.} \label{tab:xmm_obs}
   \label{tab:observations}
   \begin{threeparttable}
    \begin{tabular}{lcccc}
    \hline
    Observation ID & Revolution number & Observational date & Exposure time$^{a}$ & Count rate$^{b}$  \\
    & & & (ks) & (count s$^{-1}$)  \\
    \hline
    \multicolumn{5}{c}{\bf IRAS~13224--3809} \\ 
0673580101	&	2126	&	2011-07-19	&	34.08	&	0.37	\\
0673580201	&	2127	&	2011-07-21	&	49.26	&	0.24	\\
0673580301	&	2129	&	2011-07-25	&	52.03	&	0.09	\\
0673580401	&	2131	&	2011-07-29	&	85.11	&	0.28	\\
0780560101	&	3037	&	2016-07-08	&	39.38	&	0.19	\\
0780561301	&	3038	&	2016-07-10	&	112.35	&	0.26	\\
0780561401	&	3039	&	2016-07-12	&	99.76	&	0.22	\\
0780561501	&	3043	&	2016-07-20	&	93.22	&	0.13	\\
0780561601	&	3044	&	2016-07-22	&	97.85	&	0.38	\\
0780561701	&	3045	&	2016-07-24	&	100.13	&	0.16	\\
0792180101	&	3046	&	2016-07-26	&	110.64	&	0.13	\\
0792180201	&	3048	&	2016-07-30	&	110.55	&	0.19	\\
0792180301	&	3049	&	2016-08-01	&	86.44	&	0.07	\\
0792180401	&	3050	&	2016-08-03	&	98.63	&	0.75	\\
0792180501	&	3052	&	2016-08-07	&	102.66	&	0.23	\\
0792180601	&	3053	&	2016-08-09	&	101.50	&	0.68	\\
\hline
\multicolumn{5}{c}{\bf 1H~0707--495} \\
110890201	&	159	&	2000-10-21	&	29.94	&	0.13	\\
148010301	&	521	&	2002-10-13	&	57.98	&	0.66	\\
506200301	&	1360	&	2007-05-14	&	33.38	&	0.25	\\
506200501	&	1379	&	2007-06-20	&	20.52	&	1.03	\\
511580101	&	1491	&	2008-01-29	&	68.69	&	0.60	\\
511580201	&	1492	&	2008-01-31	&	43.88	&	1.01	\\
511580301	&	1493	&	2008-02-02	&	37.39	&	0.66	\\
511580401	&	1494	&	2008-02-04	&	50.62	&	0.56	\\
653510301	&	1971	&	2010-09-13	&	93.30	&	0.53	\\
653510401	&	1972	&	2010-09-15	&	78.46	&	0.80	\\
653510501	&	1973	&	2010-09-17	&	81.51	&	0.49	\\
653510601	&	1974	&	2010-09-19	&	79.65	&	0.57	\\
554710801	&	2032	&	2011-01-12	&	48.54	&	0.04	\\
     \hline
     \end{tabular}
    \begin{tablenotes}
    \item \textit{Note:} $^{a}$Useful exposure time used in the analysis. $^{b}$The background-subtracted count rate in the 1--4 keV energy band of the AGN obtained from the pn detector.  
    \end{tablenotes}
    \end{threeparttable}
    \end{center}
\end{table*}

We then extracted the background-subtracted light curves of the AGN in the soft (0.3–1 keV) and hard (1.2–5 keV) energy bands -- defined as reflection dominated and continuum dominated energy bands, respectively -- from the observational data flagged with {\tt PATTERN} $\leqslant$ 4 and {\tt \#XMMEA\_EP} for each observation. The corresponding PSDs were derived from the light curves which were divided into the length of $\sim$20 ks and re-binned to have a time resolution of 179 s. To improve the data signal to noise, we also re-binned the PSDs in which every bin width is equal in logarithmic space of base 1.06. These PSDs were then used as a representative of the real AGN sample in the RFR analysis in Section \ref{sec:fitting}.

\section{Reverberation and PSD model}

The variability power of the AGN on different timescales can be represented using the PSD whose intrinsic shape follows a broadband power-law profile usually simplified in the form of 
\begin{equation}
 P_{0}(f) \propto f^{- \gamma } \;,
\end{equation}
where \(f\) is the Fourier frequency and \(\gamma \)  is the power spectral index.

However, some characteristic features such as the dip and oscillatory structures caused by the X-ray reverberation can also be imprinted on the PSD profiles. We can assume that the total light curve includes both direct continuum and reverberation flux, and the total response function of the system can be written in the form of \citep[e.g.][]{Papadakis2016}
\begin{equation}
 \Psi (t) = \delta (t)+ R_{F}\psi (t) \;,
\end{equation}
where $\delta(t)$ is a delta function representing the continuum flux response and $\psi(t)$ is the disc response function representing the amount of the reflection flux detected by a distant observer, as a function of time, after the primary continuum flux. The precise $\psi(t)$ is obtained via the ray-tracing simulations \citep[e.g.][]{Cackett2014, Chainakun2016, Wilkins2016, Epitropakis2016}. Here, we use the {\sc kynxilrev} model \cite[]{Caballero2018, Caballero2020} to compute the disc response functions under the lamp-post geometry where an accretion disc is optically thick and geometrically thin and is illuminated by a point-like corona located on the rotational axis of the central black hole. The area under the disc response function is normalized to 1, so that the reflection fraction $R_{F}$ determines the ratio of (reflection flux / continuum flux). 

Based on the convolution theorem, the PSD including the X-ray reverberation effects can be obtained by \cite[]{Uttley2014,Papadakis2016, Chainakun2022}
\begin{equation}
 P_\text{obs}(f) =\left | \Psi (f) \right |^{2} P_{0}(f) \; ,
\end{equation}
where $\Psi(f)$ is the Fourier form of the total response function. The intrinsic $P_{0}(f)$ is then an input signal which is processed by the total response function resulting in a reverberation feature appearing as an oscillatory structure in the $P_\text{obs}(f)$ \citep{Papadakis2016, Chainakun2019a}.

By using the {\sc kynxilrev} model \citep{Caballero2018, Caballero2020} to produce the disc response functions, the continuum spectrum is represented by a cut-off power-law: $ F(E) \propto E^{-\Gamma } E^{-E/E_{c}}$, where $E_{c} = 300$~ keV. The disc extends from the innermost stable circular orbit (ISCO) to 1000~$r_{\rm g}$. We initially fix the black hole spin to be $a=1$. Other parameters in the {\sc kynxilrev} model, if not specified, are set to their default values.

\begin{figure*}
\centerline{
\includegraphics*[width=0.5\textwidth]{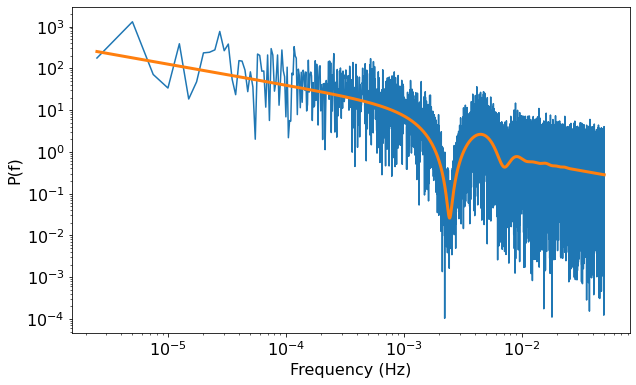}
\put(-52,135){$h = 4~r_{\rm g}$}
\includegraphics*[width=0.5\textwidth]{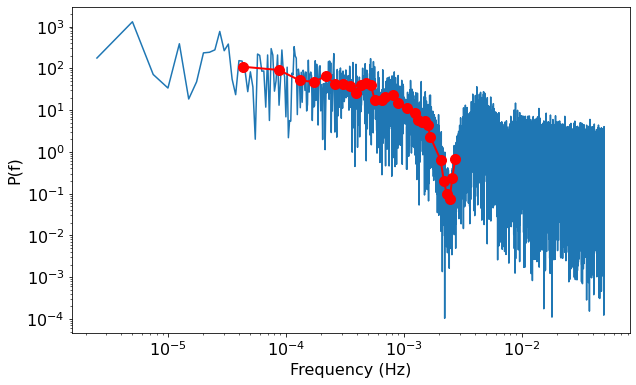}
}
\centerline{
\includegraphics[width=0.5\textwidth]{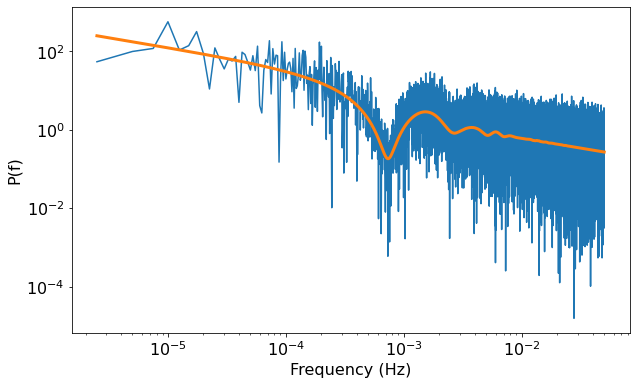}
\put(-52,135){$h = 26~r_{\rm g}$}
\includegraphics[width=0.5\textwidth]{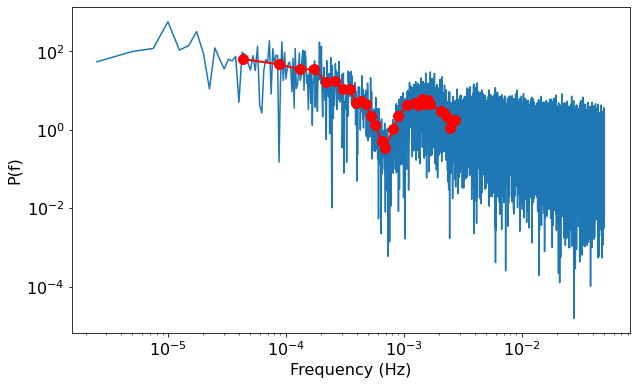}
}
\caption{Left panel: Clean PSD (orange lines) in the form of a power law and the corresponding noisy PSD (blue lines) generated using the method of \protect\cite{Timmer1995}, imprinted with the reverberation echoes with the source height $h=4~r_{\rm g}$ (top panels) and $h=26~r_{\rm g}$ (bottom panels). Right panel: Noisy PSD data after being binned and limited the frequency range to be $\sim 3 \times 10^{-5}$--$3 \times 10^{-3}$~Hz (red points), as expected from IRAS~13224--3809 observations.}
\label{fig-example}
\end{figure*}

Examples of the simulated PSD profiles are presented in Fig.~\ref{fig-example}. In this illustration, we assume the black hole mass of $10^{6}~M_{\odot}$ and the inclination $i = 45^{\circ}$. The orange lines in the left panels represent the PSD that already include the reverberation effects due to the lamp-post corona. The main reverberation dip, as expected, shifts towards higher frequencies (shorter timescales) for lower coronal height. The corresponding noisy PSD profiles (blue lines) are simulated using the \cite{Timmer1995} method, then are binned into $\sim 30$ frequency bins within the frequency range of $\sim 3 \times 10^{-5}$--$3 \times 10^{-3}$~Hz (red points in the right panel), similar to what is normally extracted from the {\it XMM-Newton} observations of IRAS~13224--3809. After being binned, these noisy PSD data are then used to train the machine, as their shapes and frequency bins are quite close to the observational ones.

\section{Machine learning algorithms}

The machine learning (ML) algorithm used here is the random forest regressor (RFR). It aims to probe the characteristic features of the reverberation signatures that appear on the PSD profiles in order to predict the disc corona geometry, such as the coronal height and the disc inclination. We follow the standard procedure by splitting the PSD data into 80/20 for training/testing. All ML algorithms and related functions for the RFR model development are adopted from {\sc scikit-learn}\footnote{\url{https://scikit-learn.org/}} \citep{Pedregosa2012}.

\subsection{Random forest regressor}

We use {\tt sklearn.ensemble.RandomForestRegressor()} to implement the RFR algorithm to the model. The RFR is an ensemble method known as the bootstrapping random forest. The RFR algorithm combines the prediction result of multiple decision trees (DTs) to obtain a final powerful model for regression. Each DT starts at the top with the root node that is split into a left and right node, which is then split further in a tree structure until the model has a satisfying result. At each node, one of the data features is evaluated so that a certain path (i.e. branch) is determined toward the final node of the tree (referred to as the leaf node) where a numerical value is best estimated.

\begin{figure}
    \centerline{
        \includegraphics[width=0.5\textwidth]{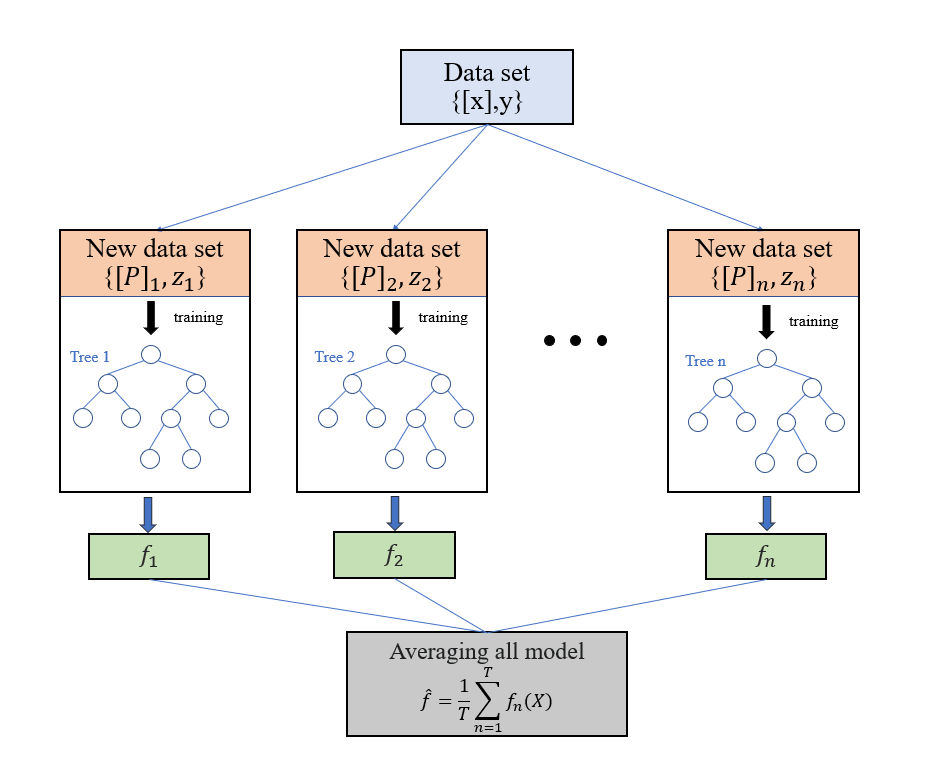}
    }
    \caption{ Flowchart of the random forest bootstrap processing, where $[x]$ is the complete PSD data set with the corresponding response $y$. $[P]_{n}$ is the new PSD data set for individual decision tree after bootstrapping process with the corresponding response $Z_{n}$. The data are trained and tested individually via the decision-tree model $f_n$ ($n=1,2,3,...,T$), where $T$ is the total number of trees in the forest. The final model is then obtained by averaging the results from all decision-tree models.  
    \label{fig-rfr}}

\end{figure}

During the bootstrapping process (Fig.~\ref{fig-rfr}), the RF algorithm randomly selects, with replacements, the PSD samples in the training set (e.g. some PSD data may be repeated in the subset while others may not be included at all). This creates a new subset with a different set of PSD profiles for one decision tree. This process is repeated $T$ times meaning that we have $T$ decision trees, $f_n$, ($n=1,2,3,...,T$) in the forest. For each individual tree, the data are trained and tested via the model given by $f_n$. The ensemble learning is initiated by using $f_n$ to predict the unseen data set $X$. After that, the result is obtained by averaging each individual prediction on unseen samples. This process leads to a final RFR model that can be written as
\begin{equation}
\hat{f}= \frac{1}{T} \sum _{n=1}^{T} f_{n}(X) \;.
\end{equation}

\subsection{Hyperparameter tuning and cross-validation}

Hyperparameters are variables that are set before the learning phase. They help control and optimize the learning algorithm and are essential for achieving an accurate model. There are two main hyperparameters for the RFR model that we select to fine-tune, which are the maximum depth of the trees and the number of trees (estimators) in the forest.

To fine-tune these hyperparameters, we use {\tt sklearn.model\_selection.GridSearchCV()} that allows us to define a grid of possible values of each hyperparameter, before training the model using each combination of these values. The model is trained and validated via the $K$-fold cross-validation process, which randomly divides the training set further into $K$ folds. The model is trained using the data in $K-1$ folds (new training set) and is evaluated using the remaining fold (validation set). The process is repeated $K$ times, with the model being trained and validated on different folds. Here, we fix $K=5$, meaning that the training data are split into 5 folds having 4 folds (80\%) for training and 1 fold (20\%) for validating. The data that are initially split into the test set are then kept unseen all the time for the final evaluation.

After the cross-validation process, the best combination of hyperparameters is identified. For the final evaluation, the developed model is used to predict the data in the test set that are kept unseen during the training phase, so they represent the data that are completely new to the machine.

\subsection{Data preparation and preprocessing}

To prepare our data set, we simulated the response functions, $\psi (t)$, using the {\sc kynxilrev} model by varying the source heights, $h$, between $2.3-30~r_{\rm g}$ and the reflection fraction, $R_F$, between $0.6-1.6$. The PSD index, $\gamma$, is varied between $0.5-2.5$. For the black hole mass and inclination, we investigate the cases when they are either fixed or free. We also investigate other scenarios such as using different binning or removing some frequency bins (e.g. those that are dominated by Poisson noises) when we train and test the machine. These aim to determine the best way of data processing for the model to accurately predict the coronal height from real observational PSD data.

\section{Results and analysis}

\subsection{Efficiency of RFR algorithm}

\begin{figure}
    \centerline{
        \includegraphics[width=0.45\textwidth]{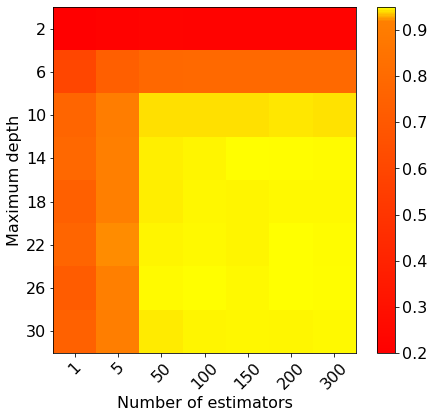}
        \put(-29,23){$R^2$}
    }
    \caption{Cross validation score ($R^2$) for predicting $h$ for different values of hyperparameters. The highest $R^{2}_{\rm train}$ is 0.96, corresponding to the case when the number of estimators is 200 and the maximum depth is 26. 
    \label{fig-cv}}
\end{figure}

Our simulated PSD profiles in the training data set are used for fine-tuning the hyperparameters via 5-fold cross-validation. Fig.~\ref{fig-cv} shows the accuracy scores for predicting the coronal height obtained from the training/validation process. In this illustration, we fix the black hole mass $M=2 \times 10^{6}~M_{\odot}$, the inclination $i=45^{\circ}$ and use the number of frequency bins $N_{\rm bin}=30$. The RFR model can accurately predict the coronal height when the number of estimators is $\gtrsim 50$ and the maximum tree depth is $\gtrsim 10$, with the highest accuracy score of $R^{2}_{\rm train}$ $\sim 0.96$. The best number of estimators and maximum depth obtained here are 200 and 26, respectively. Note that the $R^2$ score\footnote{More explanation on the $R^2$ score at \url{https://scikit-learn.org/stable/modules/generated/sklearn.metrics.r2_score.html}} is the proportion of the variation in the dependent variable that is predictable from the independent variable. It is usually used to evaluate the ML model's performance. The maximum $R^2$ score is 1 which means that the model fits the data perfectly, or all independent variables can explain all possible interpretations of the dependent variable.

\begin{figure}
    \centerline{
        \includegraphics[width=0.5\textwidth]{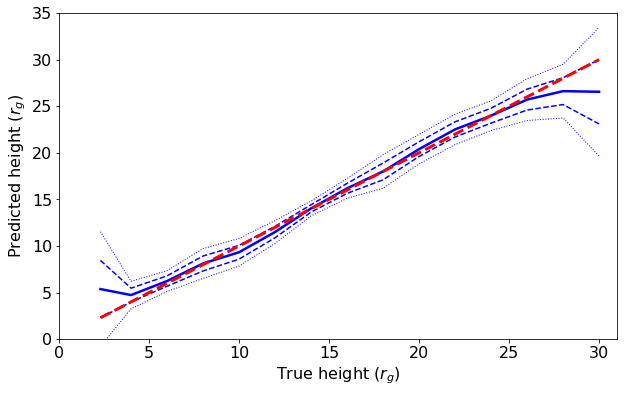}
    }
    \caption{Plot of the true versus predicted values of the coronal height using the test data set. The $R^{2}_{\rm test}$ is 0.95, which is comparable to the $R^{2}_{\rm train}$. The blue solid line represents the results obtained from the RFR model. The red dashed line shows the perfect prediction line. The regions between the blue-dashed lines and between the blue-dotted lines represent the deviations of the predicted heights within $\pm 1\sigma$ and $\pm 2\sigma$, respectively.
    \label{fig-h}}

\end{figure}

Then, the developed RFR model is tested with the new, unseen data spared in the test data set. The true versus predicted values of the coronal height in this case is shown in Fig.~\ref{fig-h}. The obtained $R^{2}_{\rm test}$ is 0.95. Since $R^{2}_{\rm test} \sim R^{2}_{\rm train}$, the model is generalized well (i.e. not overfitting the data). However, the model accuracy is relatively small at the low and high ends for the source height. The low accuracy at $h \sim 2.3 \ r_{\rm g}$ is because, with this assumed mass, the reverberation features are clearly noticeable at the frequencies of $\gtrsim 3 \times 10^{-3}$~Hz, or beyond the highest frequency probed in the PSD profiles. For $h \gtrsim 28 \ r_{\rm g}$, less accuracy may arise because the reverberation bump is too weak in the majority of the PSD data, especially when the reflection fraction is also low.

\begin{figure}
    \centerline{
        \includegraphics[width=0.5\textwidth]{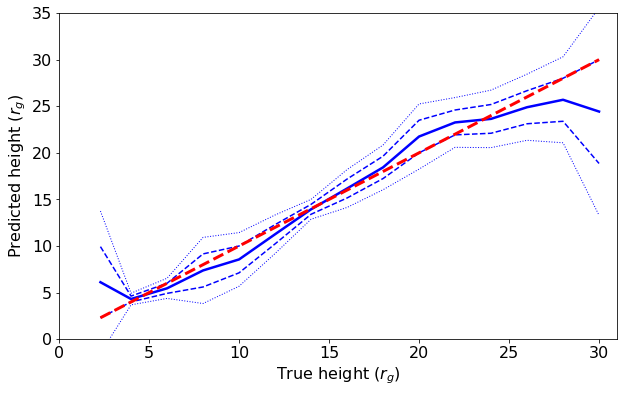}
        \put(-210,140){$R_{F} < 1$}
    }
    \centerline{
        \includegraphics[width=0.5\textwidth]{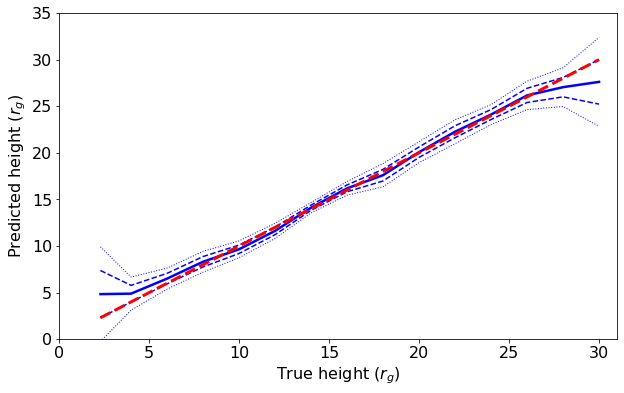}
        \put(-210,140){$R_{F} \geq 1$}
    }
    \caption{Same as in Fig.~\ref{fig-h}, but showing the separate cases when the reflection fraction $R_{F}<1$ (top panel) and $R_{F} \geq 1$ (bottom panel).}
    \label{fig-R}
\end{figure}

Furthermore, Fig.~\ref{fig-R} shows that the RFR model is more effective in cases of the reflection fraction $R_{F} \geq 1$ ($R^{2}_{\rm test}=0.98$). This means that it is best to apply the model to the observational data extracted in the reverberation dominated band (e.g. $0.3 - 1$~keV band).

When the model is allowed to predict both coronal height and inclination, the accuracy is relatively small ($R^{2}_{\rm test} < 0.5$). This suggests that the model can make an accurate prediction for the coronal height, but not for the inclination. The poor prediction also occurs when we try to predict both source height and black hole mass, even when the reflection fraction $R_{F}>1$. Making an accurate prediction for all these parameters simultaneously may require fine-tuning more hyperparameters, so at this stage we select to fix $i$ and $M$, and investigate further the effects of the data binning. There are also intrinsic degeneracies between model parameters (see discussion Section~\ref{sec:discussion}).

\begin{figure}
    \centerline{
        \includegraphics[width=0.5\textwidth]{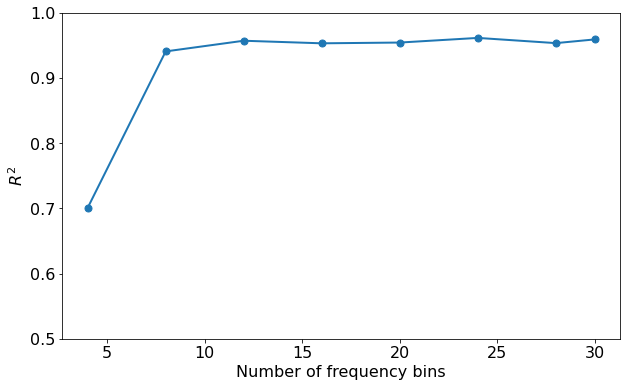}
    }
    \centerline{
    \hspace{0.45cm}
        \includegraphics[width=0.53\textwidth]{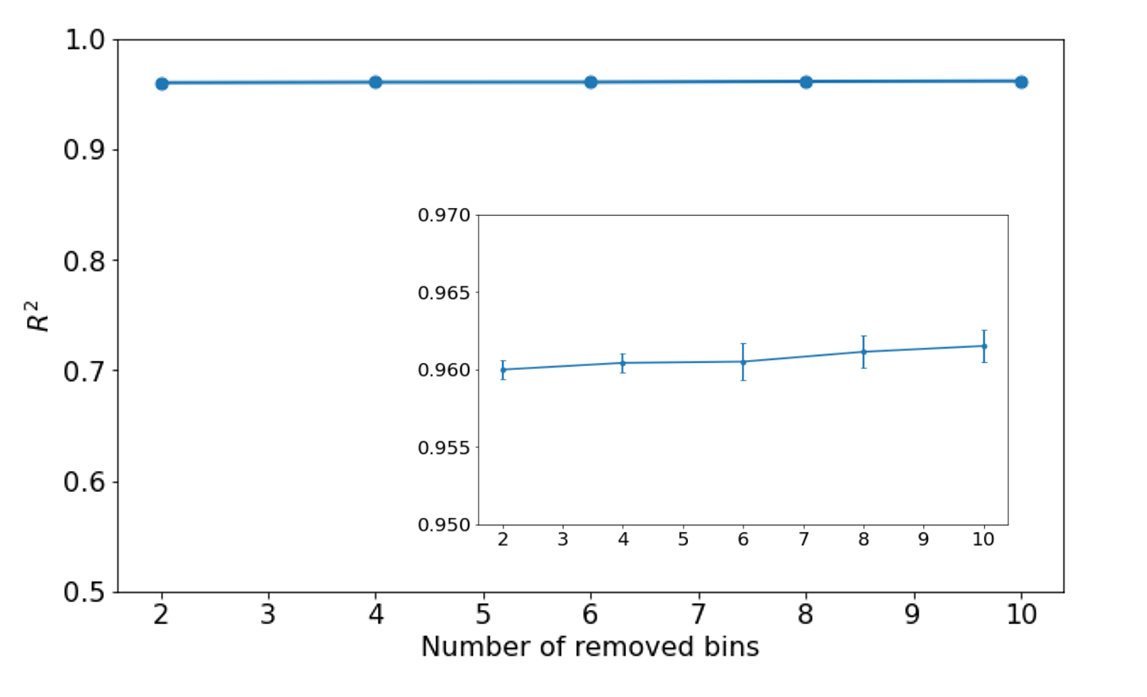}
    }
    \caption{
    Accuracy score ($R^{2}_{\rm test}$) for predicting $h$ varying with the number of frequency bins, $N_{\rm bin}$, when the frequency range of the PSD is fixed at $\sim 10^{-5}$--$10^{-3}$~Hz (top panel). We also show the $R^{2}_{\rm test}$ profile when we fix $N_{\rm bin} = 30$ before some of the frequency bins are randomly removed (bottom panel). The sub-figure highlights the error bars obtained by repeating the random process 20 times. This illustrates that the $R^{2}_{\rm test}$ is still $\gtrsim 0.95$ even if the observational data are pre-processed by removing the Poission-noise dominated bins, as long as the machine is trained using the simulated PSD profiles that are binned consistently with the observational data. Note that $R^{2}_{\rm train} \sim R^{2}_{\rm test}$, so the $R^{2}_{\rm train}$ profile is omitted for clarity of the plots.}
    \label{fig-bin}
\end{figure}

Fig.~\ref{fig-bin} (top panel) shows how the number of frequency bins ($N_{\rm bin}$) affects the $R^{2}_{\rm test}$ when we fix $M = 2 \times 10^{6}~M_{\odot}$, $i=45^{\circ}$ and produce the PSD in the frequency range of $\sim 10^{-5}$--$10^{-3}$~Hz. The developed RFR model seems to work very well ($R^{2}_{\rm test} \gtrsim 0.9$) when $N_{\rm bin} \gtrsim 8$. The observed PSD data of IRAS 13224--3809 and 1H~0707--495 are extracted using the number of frequency bins of $\sim 30$, which is definitely large enough that the reverberation features can still be interpretable by the model. Nevertheless, some frequency bins of the observational data can be dominated by Poisson noise. In this case, we investigate whether it will affect the accuracy score if we remove those Poisson-dominated bins. This is illustrated in Fig.~\ref{fig-bin} (bottom panel) when $2-10$ bins are randomly removed from the simulated PSD profiles, and the machine is trained and tested with these modified data. The $R^{2}_{\rm test} \gtrsim 0.95$ in all cases suggests that the observational PSD data can be pre-processed by removing the Poission-noise dominated bins, as long as the machine is trained using the simulated PSD profiles that are modified and binned in a similar way.

\begin{figure}
    \centerline{
        \includegraphics[width=0.5\textwidth]{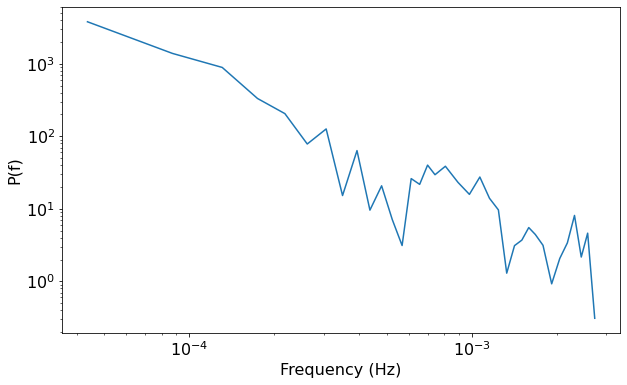}
\put(-81,134){IRAS~13224--3809}
\put(-71,124){(Rev.~no.~2127)}
    }
    \centerline{
        \includegraphics[width=0.5\textwidth]{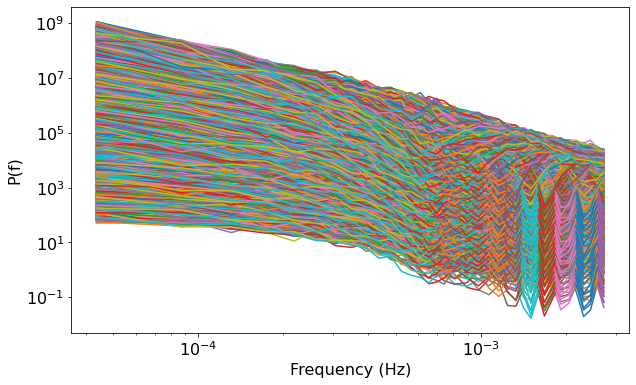}
        \put(-88,134){Simulatated profiles}
    }
    \caption{Top panel: Example of the observed PSD profile of IRAS~13224--3809 (Rev.~no.~2127) extracted in the 0.3--1 keV. The bins where the Poisson noise dominate are removed. Bottom panel: Simulated corresponding PSD profiles with the mass and inclination fixed at $M = 2 \times 10^{6}~M_{\odot}$ and $i=45^{\circ}$, as expected for IRAS~13224--3809. The simulated profiles are binned specifically to Rev.~no.~2127. Some frequency bins are removed similar to what is done to the observed PSD data.}
    \label{fig-PSD-obs-ML}
\end{figure}

Fig.~\ref{fig-PSD-obs-ML} shows examples of the observational PSD data of IRAS~13224--3809 (Rev. no. 2127) and the corresponding PSD profiles simulated for training and testing the machine. Their bins, as well as frequency range, are adjusted to be the same. The frequency bins of the observed PSD that are dominated by Poisson noise are removed, and these bins are removed from the simulated data too. Therefore, to predict the coronal height from each observation, each RFR model must be newly developed using the simulated PSD profiles that are specific to each individual observation. In this way, we can always ensure, based on the performance of the model, that the predicted coronal height is highly accurate ($R^{2} \gtrsim 0.9$).

\subsection{Fitting IRAS~13224--3809 and 1H~0707--495}
\label{sec:fitting}

\begin{figure}
    \centerline{
        \includegraphics[width=0.5\textwidth]{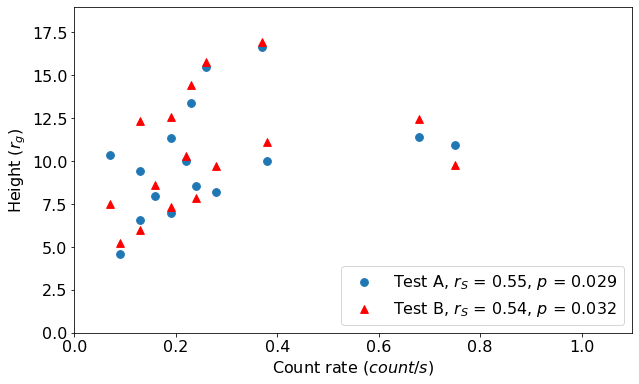}
        \put(-80,134){IRAS~13224--3809}
    }
    \centerline{
        \includegraphics[width=0.5\textwidth]{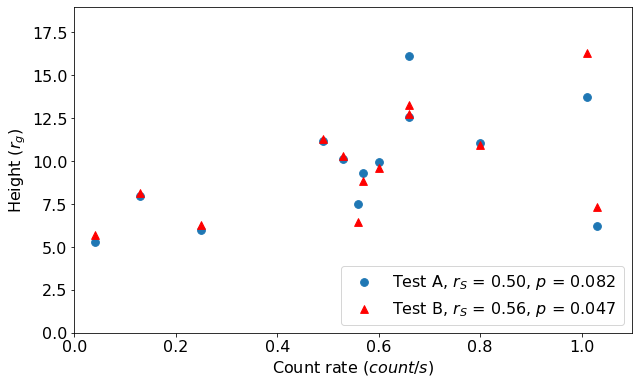}
        \put(-60,134){1H~0707--495}
    }
    \caption{Coronal height, $h$, versus count rate of the observations of IRAS~13224--3809 (upper panel) and 1H~0707--495 (lower panel) as predicted by the RFR models when using the PSD extracted in 0.3--1~keV band. Test A and B refer to the cases when the power index of the training PSD data is allowed to vary between 0.5--2.5 and 0.5--4.0, respectively. The Spearman correlation coefficient ($r_S$) and $p$-value ($p$) are also presented.
    \label{fig-observed-data}}
    \end{figure}

For IRAS~13224--3809, we fix $M=2\times10^{6}~M_{\odot}$ \citep{Alston2020} and $i=45^{\circ}$ \citep{Caballero2020}. Meanwhile, for 1H~0707--495, we fix $M=2.3\times10^{6}~M_{\odot}$ \citep{Zhou2005} and $i=45^{\circ}$, which is an intermediate value between what's reported in \cite{Caballero2018} and \cite{Wilkins2011}. The developed RFR algorithm is then applied to predict the coronal height in 16 and 13 observations of IRAS~13224--3809 and 1H~0707--495, respectively, where their PSDs are extracted in the 0.3--1~keV band. We input each individual observed PSD data, one by one, and the specific RFR model can be produced. We find both $R^{2}_{\rm train}$ and $R^{2}_{\rm test}$ are comparable, which are $\gtrsim 0.9$ for all observational data. The predicted coronal heights versus the count rates for both AGN are presented in Fig.~\ref{fig-observed-data}. We find a moderate, monotonic correlation between coronal height and count rate in both IRAS~13224--3809 and 1H~0707--495. The mean squared errors are varied between $\sim 0.2-2.8$ among different data sets. We investigate two further cases, referred to as Test A and Test B, where we sample 20 PSD indices in the range of 0.5--2 and 40 PSD indices in the wider range of 0.5--4, respectively. We find that the predicted source heights are slightly different if we assume different ranges of the power law PSD indices. This, however, does not change the trend of increasing source height with count rate.

\begin{figure}
    \centerline{

        \includegraphics[width=0.5\textwidth]{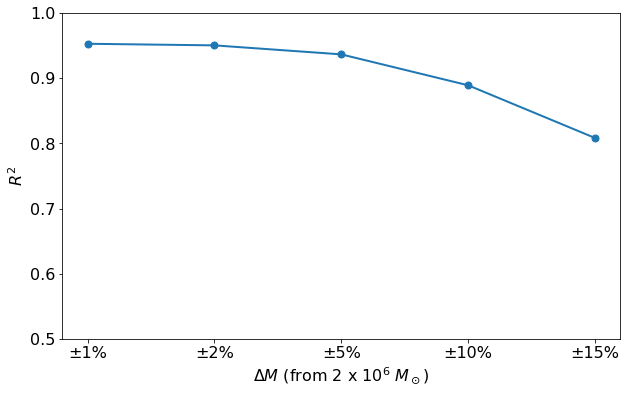}
    }
    \caption{Accuracy score ($R^{2}_{\rm test}$) for predicting $h$ when the RFR model is trained with the data where the mass is fixed at $2 \times 10^{6}~M_{\odot}$, but is tested with new data that have different values of the black hole mass. In this illustration, we use the frequency range of $\sim 10^{-5}$--$10^{-3}$~Hz and $N_{\rm bin} = 30$. For clarity, the $R^{2}_{\rm train}$ profile is omitted since $R^{2}_{\rm train} \sim R^{2}_{\rm test}$.
    \label{fig-M}}
\end{figure}

Fig.~\ref{fig-M} shows an example of how the obtained accuracy changes if the mass of the target AGN differs from the value used to train the machine. If the mass difference is larger, the accuracy is undoubtedly smaller.  We find that the $R^{2} \gtrsim 0.9$ can still be achieved if the true mass is $\pm10\% $ from the trained value.

\section{Discussion and Conclusion}
\label{sec:discussion}
Firstly, we investigate the effectiveness of the RFR model in estimating three physical parameters of AGN including the source height, inclination, and black hole mass. Our results reveal that the accuracy for forecasting the inclination is poor ($R^{2}<0.5$) because either the reverberation feature is indistinct to be extracted or the model requires a larger number of PSDs varying with inclination to be trained. Similarly, the accuracy in determining the black hole mass simultaneously with the coronal height is poor as the characteristics of PSDs varying with the mass can resemble those changing with the source height \citep{Papadakis2016}. For example, the data with smaller mass and larger height can also be reasonably explained by the models with larger mass but smaller height. The degeneracy between the mass and the source height is also found in other timing profiles such as the lag-frequency spectrum \citep[e.g.][]{Cackett2014, Caballero2018, Caballero2020}, and still cannot be easily broken by the model. 

In the case of predicting the source height alone, the obtained accuracy is excellent ($R^{2} \gtrsim 0.9$), particularly when the reflection fraction $R_{F} \geqslant 1$. This is because the reverberation feature in the PSD, such as the dip, is more pronounced. As a result, we focus on the PSD data extracted in the energy band of $0.3$--1~keV where the reflection flux is usually dominated and choose to predict the source height by fixing the black hole mass and the inclination angle.

\cite{Chainakun2021} proved that the ML technique can be used to extract the reverberation signatures on the PSD profiles to predict the source height. However, for the ML approach, there is for a classification problem where the source heights are categorized into several distinct classes (i.e. discrete values), each of which represents a range of source heights. Here, we use the RFR algorithms so that the source height can be predicted as continuous values. The simulated PSD profiles being used to train, cross-validate and test the machine are also specifically produced using a similar frequency range and a similar number of frequency bins to each observational data. Higher Poisson noise can lead to higher standard deviation of the measured lag, without affecting the lag value \citep{Kara2013b}. Rather than quantifying the Poisson noise level in each observation and involving the errors from the PSD measurement, we select to directly remove the Poisson-noise dominated bins instead. This is acceptable as long as the training and test data are processed in the same way. The prediction accuracy of $R^{2} > 0.9$ can still be achieved if the number of Poisson-noise dominated bins being removed is less than 10, which is likely the case for our observational data.

Once the developed RFR model is applied to observational data, we find the IRAS~13224–3809 corona varies between $\sim 5-18~r_{\rm g}$. This is comparable with the constrained coronal height of $\sim 6-20~r_{\rm g}$ using the lag-frequency spectra \citep{Alston2020}. The combined spectral-timing analysis by \cite{Caballero2020} also reveals the coronal height varying between $\sim 3-10^{+10}_{-1}~r_{\rm g}$, assuming a maximally spinning black hole. Recently, tracing the corona evolution from the reverberation signatures that appear in the PSD of IRAS~13224–3809, but with conventional fitting technique, suggests the varying corona located at the height of $\sim 3-25~r_{\rm g}$ \citep{Chainakun2022}. We also find a moderate monotonic correlation showing the trend of increasing source height with the count rate, $r_{S} \sim 0.55$ ($p \sim 0.03$), similar to \cite{Alston2020, Caballero2020, Chainakun2022}. This is because the photons emitted from the corona located further away from the black hole can escape the strong gravitational field more easily than those emitted from the corona closer to the black hole, leading to higher source luminosity. The corona evolution in IRAS~13224–3809 then is quite consistent among these studies, even though different methods are used. 

Here, we fix the IRAS~13224–3809 mass to be $2 \times 10^{6}~M_{\odot}$ \citep{Alston2020} and employ the intermediate inclination of $45^{\circ}$ \citep{Caballero2020}. The result suggests that high accuracy can still be obtained as long as the fixed mass is $\lesssim 10$ per cent different from the true value. The spectral fits by \cite{Jiang2022} using a high-density disc model suggested the inclination of $60^{\circ}$--$70^{\circ}$. The choice of inclination, however, should not significantly affect the accuracy in predicting the coronal height since the reverberation features in timing profiles are less dependent with the inclination \citep[e.g.][]{Cackett2014, Papadakis2016}. Recently, \cite{Hancock2023} fitted an extended corona model, simplified using dual lamp-post sources, to the lag-frequency spectra of IRAS~13224–3809. By adopting the inclination $i=64^{\circ}$ as reported by \cite{Fabian2013} and placing the first, lower source at $2~r_{\rm g}$, they found that the second, upper source could vary between $\sim 3-20~r_{\rm g}$. Although different coronal geometry and inclination are assumed, the disc-corona distance constrained here is comparable with the upper source height suggested in \cite{Hancock2023}. Developing the RFR model to probe the spatial extent of the corona is possible, but it is a subject for future research. 

Recently, \cite{Chainakun2023} employed the Granger-causality test to probe the intrinsic reverberation lags in IRAS 13224–3809 to be $\sim 200$--500~s. They directly converted the intrinsic lags to the true light-travel distance of $\sim 20$--50~$r_{\rm g}$, and by assuming a face-on disc the coronal height could be roughly approximated to be $\sim10$--25~$r_{\rm g}$. Furthermore, they reported the evidence of the corona variability within some individual observations (e.g. the corona can change from $\sim10$--25~$r_{\rm g}$ to $\sim 55$~$r_{\rm g}$, or vice versa, from the beginning to the end of the observation). The method outlined in \cite{Chainakun2023} is a statistical test and the large coronal height (e.g. $> 25$~$r_{\rm g}$) is seen only in some partial segments of the light curves. The different lags from different segments can lead to an insignificance of the lags at a particular value when analysing the full light curve, or the multiple lags can be averaged out. However, \cite{Chainakun2023} still found, for the majority of the IRAS 13224–3809 observations, that the corona situated at $\sim10$--25~$r_{\rm g}$, so with the measurement error of $\sim 5$~$r_{\rm g}$ suggested by their method, the coronal height constrained there is mostly comparable to what obtained in this work.

For 1H~0707–495, the RFR model places the corona at $h \sim 5-18~r_{\rm g}$ as well. The hint of increasing source height with count rate is also observed. Moreover, \cite{Chainakun2021} suggested that even if the ML is trained using the power-law PSDs but the observed data have a bending power-law shape, the coronal height can still be forecasted with the accuracy of $\gtrsim 0.75$. Then, the obtained accuracy reported here can be lower if the intrinsic shape of the PSD, which is not known prior, has a bending power-law form. Perhaps, the best way might be to train the machine with the PSD with various intrinsic shapes as much as possible, but this will progressively increase computational time consuming and is beyond the scope of this work. Also, when the range and number of the photon indices are changed, the regression results show slight modifications (see Test A and Test B in Fig.~\ref{fig-observed-data}). This effect, however, should not change the overall trend of the correlation if the range of the photon indices are large enough. 

Nevertheless, the nature of 1H~0707–495 is quite uncertain. Some suggested that it should have a low-spin black hole \citep{Done2016}, while the mass can possibly be $\sim 5-10 \times 10^{6}~M_{\odot}$ \citep[e.g.][]{Zoghbi2011,Pan2016}, which is larger than the mass assumed in, e.g., this work and \cite{Szanecki2020} by a factor of $\sim 2$--5. \cite{Szanecki2020} analysed the relativistic reflection spectra from the disc around a maximally spinning black hole and found the corona is very compact, with an extended size no larger than $\sim 1~r_{\rm g}$ and located at $h \sim 3~r_{\rm g}$. A compact corona at $\sim 3$--$4~r_{\rm g}$ was also proposed by \cite{Dauser2012} and \cite{Caballero2018}. Contrarily, \cite{Wilkins2012} studied the emissivity profiles due to various coronal geometries  along with previous constraints from reverberation lags, and inferred a presence of the 1H~0707--495 coronal height at $\sim 2~r_{\rm g}$ that radially extends outwards to a radius of  $\sim 30~r_{\rm g}$. The corona also seems to expand as the source luminosity increases \citep{Wilkins2014}. \cite{Dovciak2016} suggested that the 1H~0707–495 corona must be large enough to intercept sufficient disc photons to produce the hard X-ray continuum and, during the maximum flux, it could be at $\sim 2-5~r_{\rm g}$ extending further outside to $\sim 20~r_{\rm g}$. Furthermore, \cite{Chainakun2019b} developed a spherical corona model to explain the frequency-dependent time lags of AGN, and found that the 1H~0707--495 corona could be radially extended up to $\sim 10~r_{\rm g}$. Previous study using a dual lamp-post model also suggested the 1H 0707--495 corona could extend upwards, up to $\sim 20~r_{\rm g}$ \citep{Hancock2023}. Although this work assumes a lamp-post geometry, it suggests a dynamic corona of 1H~0707--495 that can move far beyond a few gravitational radii from the central black hole, and an increase of the coronal height with luminosity is also evidenced.

In conclusion, our study highlights the potential of machine learning, specifically regression techniques, for probing the coronal evolution in AGN. The constrained $h \sim 5-18~r_{\rm g}$ in both AGN here supports the height-changing corona and the light-bending scenario \citep{Miniutti2004}, in which the coronal height is correlated to source luminosity. The fluctuations in the gravitational and magnetic fields caused by the black hole and the accretion disc can induce the corona to move closer or farther away from the black hole, leading to variations in the X-ray emission. The results are also in line with recent studies on other AGN such as NGC~5548 that, by probing the UV/optical PSD variation, revealed a dynamic corona chainging its position and probably also its spatial extension on various timescales \citep{Panagiotou2022}.

\section*{Acknowledgements}
We thank the anonymous referee for comments that improved the paper. This research has received funding support from the NSRF via the Program Management Unit for Human Resources \& Institutional Development, Research and Innovation (grant number B16F640076). N.M. thanks Suranaree University of Technology (SUT) and National Astronomical Research Institute of Thailand (NARIT) for the financial support.

\section*{Data availability}
The observational data were accessed from {\it XMM-Newton} Observatory (\url{http://nxsa.esac.esa.int}). The response function data analysed here were generated using the {\sc kynxilrev} model available in \url{https://projects.asu.cas.cz/stronggravity/kynreverb}. The code for the developed RFR model to predict the coronal height is available in \url{https://github.com/PChainakun/RanForPSD}. Other derived data underlying this article will be shared on reasonable request to the corresponding author.


\bibliographystyle{mnras}

\bsp	
\label{lastpage}
\end{document}